\documentstyle[romp31]{article}

\title{Description of isolated macroscopic systems inside
quantum mechanics}

\author{L.~Lanz${}^1$, O.~Melsheimer${}^2$ and
B.~Vacchini${}^{1,2}$\thanks{\ \ Supported by Alexander von
Humboldt-Stiftung}
\\
${}^1$
Dipartimento di Fisica
dell'Universit\`a di Milano and INFN,
Sezione di Milano,
\\
Via Celoria 16, I--20133, Milan, Italy
\\
${}^2$
Fachbereich Physik, Philipps-Universit\"at,
Renthof 7, D--35032, Marburg, Germany
\\
(e-mail:
lanz@mi.infn.it;
melsheim@mailer.uni-marburg.de;
vacchini@mi.infn.it)
\\[2ex]
}

\begin{document}

\hyphenation{
non-equi-li-bri-um
}

\maketitle
\begin{abstract}
For an isolated macrosystem classical state parameters
$\zeta(t)$ are introduced inside a quantum mechanical
treatment. By a suitable mathematical representation of the
actual preparation procedure in the time interval
$[T,t_0]$
a statistical operator is constructed as a solution of the
Liouville von Neumann equation, exhibiting at time $t$ the
state parameters $\zeta(t')$, $t_0\leq t' \leq t$, and
{\it  preparation  parameters} related to times
$T \leq t'\leq t_0$. Relation with Zubarev's non-equilibrium
statistical operator is discussed. A mechanism for memory
loss is investigated and time evolution by a semigroup
is obtained for a restricted set of relevant observables,
slowly varying on a suitable time scale.
\end{abstract}

\section{Introduction}
Quantum mechanical non-separability is generally considered
as an obstacle for an objective description of physical
systems. Such obstacle can be partly overcome if one consistently
takes quantum mechanics as a description of preparation and
measuring procedures, rather than of the intrinsic structure
of things.
The assignment of a suitable statistical operator to
represent an objectively given preparation  procedure is
therefore the turning point: that this is not a simple task
is immediately clear if one realizes that this must be done
inside a framework in which at least isolated macrosystems
can be described; this already requires a field description
inside a Fock-space. Actually the typical parameters one
takes into account when determining a concrete experimental
realization have no direct connection with properties
related to microphysical structure of the system one is
dealing with.
Well-known examples of such  parameters are velocity,
temperature and chemical potential fields, by which a large
variety of preparations can be described; these preparations
contribute however only a very small part of those
performed using what nature already has prepared. Usually
the duration of the previously indicated preparations
referring to rather {\it simple systems} is very small,
while for complex systems it
can be extremely long. It seems reasonable to expect that the
use of any given quantum field theory is effective only if
preparations are considered of not exceedingly high
complexity.
We stress that these preparation parameters characterize
the subset of relevant statistical operators, therefore they
can be associated with each element of the prepared
statistical collection. In this way one recovers non-statistical 
features of experimental settings and then
objectivity appears.
Obviously statistics plays its role when measurements
are done on the prepared system (then it is no longer
isolated). Let us remark that also the set of meaningful
measurements is restricted and related to the  preparation
procedure, e.g., the dynamics of the relevant variables must
be {\it slow enough}; otherwise the isolation carried out
during the preparation would no longer be sufficient to
avoid influence from the environment.
Isolation of the system has a decisive role, subsets of
statistical operators must be chosen and too fast
observables must be avoided, all this amounts to the fact
that the mathematical framework must be less rigid than it
usually is inside quantum field theory in the thermodynamic
limit.

\section{Classical state parameters for a macroscopic system}
According to the general point of view described in \S~1,
we have to choose a suitable set of relevant observables.
Let the microphysical structure underlying the system be
described in terms of non-relativistic interacting
particles, associated with Schr\"odinger fields
${\hat \psi}_{\alpha}({\mbox{\bf x}},\omega)$
        \[
        {
        \left[  
        {\hat \psi}_{\alpha}({\mbox{\bf x}},\omega),
        {\hat
        \psi}_{\alpha'}^{\scriptscriptstyle\dagger}({\mbox{\bf
        x}}',\omega')
        \right]  
        }_\pm  
        =  
        \delta_{\alpha,\alpha'}
        \delta_{\omega,\omega'}
        \delta^3 ({\mbox{\bf x}}-{\mbox{\bf x}}')
        \quad
        \omega=1,2,\ldots,2s_\alpha+1
        \]
$\alpha$  denoting different types of particles and $s_\alpha$ the
corresponding spin.
If the system is confined inside a space region $\Omega$,
with boundary $\partial\Omega$, one represents the fields in
terms of a complete orthonormal set of functions (normal
modes) $u_n({\mbox{\bf x}},\omega)$, such that:
        \[
        -{
        \hbar^2  
        \over  
        2m_{\alpha}  
        }  
        \Delta_2  
        u_n({\mbox{\bf x}},\omega)= E_n
        u_n({\mbox{\bf x}},\omega) \quad {\mbox{\bf x}}\in\Omega,
        \qquad  
        u_n({\mbox{\bf x}},\omega)=0 \quad {\mbox{\bf x}}\in
        \partial\Omega
        ,
        \]
so that
$
        {\hat \psi}_{\alpha}({\mbox{\bf x}},\omega)
        =
        \sum_n
        {\hat a}_{\alpha n}
        u_n({\mbox{\bf x}},\omega)
$,
with
$
        {
        [  
        {\hat a}_{\alpha n},
        {\hat a}_{\alpha' n'}^{\scriptscriptstyle\dagger}
        ]  
        }_\pm  
        =  
        \delta_{\alpha,\alpha'}
        \delta_{n,n'}
$,
$
        {
        [  
        {\hat a}_{\alpha n},
        {\hat a}_{\alpha' n'}
        ]  
        }_\pm  
        =  
        0    
$.
The relevant variables are functions of the fields,
typically densities of conserved quantities in suitable
configuration spaces, for example:
        \begin{eqnarray*}
        &&  
        {\hat \rho}_{\alpha}({\mbox{\bf x}},\omega)
        =  
        m_\alpha
        {\hat \psi}_{\alpha}^{\scriptscriptstyle\dagger}({\mbox{\bf
        x}},\omega)
        {\hat \psi}_{\alpha}({\mbox{\bf x}},\omega),
        \quad
        {\hat {\mbox{\bf p}}}_\alpha({\mbox{\bf x}},\omega)
        =  
        \frac {1}{2}     \!  
        \left \{  
        \left[
        i\hbar
        \nabla
        {\hat
        \psi}_\alpha^{\scriptscriptstyle\dagger}({\mbox{\bf
        x}},\omega)
        \right]  
        {\hat \psi}_\alpha({\mbox{\bf x}},\omega)
        +
        h.c.
        \right \},  
        \\
        &&
        {\hat f}_\alpha({\mbox{\bf r}},{\mbox{\bf p}})
        =
        \sum_{\omega}
        \int_{\Omega} d^3\!
        {\mbox{\bf x}}
        \int_{\Omega} d^3\!
        {\mbox{\bf x}}'
        \,  
        {\hat \psi}_{\alpha}^{\scriptscriptstyle\dagger}({\mbox{\bf
        x}},\omega)
        \langle
        {\mbox{\bf x}},\omega
        |   {\hat {\mbox{\sf f}}}
        ({\mbox{\bf r}},{\mbox{\bf p}})  |
        {\mbox{\bf x}}',\omega
        \rangle
        {\hat \psi}_{\alpha}({\mbox{\bf x}}',\omega)
        \nonumber
        \end{eqnarray*}
for each kind of particle being respectively the mass
density, the momentum density, the {\it one-particle
distribution} function (${\hat {\mbox{\sf f}}}
({\mbox{\bf r}},{\mbox{\bf p}}) $ is a suitable one-particle
phase space density~\cite{Lanz5}). Also an energy density
$
{\hat e}({\mbox{\bf x}})
$
can be introduced by slightly more complicated expressions,
related to the Hamiltonian ${\hat H}$ by
        $
        {\hat H}=
        \int_{\Omega} d^3\!
        {\mbox{\bf x}}
        \,
        {\hat e}({\mbox{\bf x}})
        $.
Generally indicating these densities by
${\hat A}_j({\mbox{{\boldmath$\xi$}}})$, conservation equations
hold of the form:
$
        {\dot {{\hat A}_j}} ({\mbox{{\boldmath$\xi$}}},t)
        =
        - div{\hat {\bf J}}_j({\mbox{{\boldmath$\xi$}}},t) 
$,
where time dependence is given in Heisenberg picture
$
        {\hat A}_j({\mbox{{\boldmath$\xi$}}},t)
        =
        e^{+{{
        i
        \over
         \hbar
        }}{\hat H}t}
        {\hat A}_j({\mbox{{\boldmath$\xi$}}})
        e^{-{{
        i
        \over
         \hbar
        }}{\hat H}t}
$.
The relevance of densities of conserved quantities has
been stressed in non-equilibrium statistical mechanics,
e.g., by Zubarev~\cite{Zubarev}. They provide in most
natural way slow enough quantities, 
if smeared with sufficiently homogeneous probe functions.
A central role has the determination of a  statistical
operator  once the expectation values of the set ${\cal M}$
of relevant observables are given: in general ${\cal M}$ is
not large enough to uniquely determine a statistical
operator at any time $t$; any set
$\{
\langle
{\hat A}_j({\mbox{{\boldmath$\xi$}}})
\rangle_t
\}$
of expectations of the relevant, linearly independent
observables determines a set of statistical operators
$M_t
(
\{
\langle
{\hat A}_j({\mbox{{\boldmath$\xi$}}})
\rangle_t
\}
)
$, that we shall call {\it macrostate}~\cite{Robin}.
Inside a macrostate the criterion of maximal von Neumann
entropy (it will be assumed finite for each macrostate)
allows under very general
conditions on ${\cal M}$ to determine a unique trace class
operator having the
typical structure of Gibbs state. Let us indicate by
${\hat w}_{\zeta(t)}$
such (generalized) Gibbs state:
        \[
        {\hat w}_{\zeta(t)}
        =
        {  
        \exp
        \left \{
        {-{
        \sum_j
        \int d {\mbox{{\boldmath$\xi$}}} \,
        {\zeta}_j({\mbox{{\boldmath$\xi$}}},t)
        {\hat A}_j({\mbox{{\boldmath$\xi$}}})
        }}  
        \right \}
        \over  
        Z
        \left[
        {\zeta}_j(t)
        \right]  
        }
        \equiv
        \exp
        \left \{
        {
        -\zeta_0(t) {\hat {\bf 1}}
        -{
        \sum_j
        \int d {\mbox{{\boldmath$\xi$}}} \,
        {\zeta}_j({\mbox{{\boldmath$\xi$}}},t)
        {\hat A}_j({\mbox{{\boldmath$\xi$}}})
        }}  
        \right \}
        \]
with
$
        \zeta_0(t)=\log
        Z
        \left[
        {\zeta}_j(t)
        \right]
$,
$
        Z
        \left[
        {\zeta}_j(t)
        \right]  
        =
        {\mbox{{\rm Tr}}} \,
        {  
        \exp
        \left \{
        {-{
        \sum_j
        \int d {\mbox{{\boldmath$\xi$}}} \,
        {\zeta}_j({\mbox{{\boldmath$\xi$}}},t)
        {\hat A}_j({\mbox{{\boldmath$\xi$}}})
        }}  
        \right \}
        }
$
being the partition function of the system at time $t$,
while ${\zeta}_j({\mbox{{\boldmath$\xi$}}},t)$
are the Lagrange parameters related to the maximization
procedure.
${\hat w}_{\zeta(t)}$ represents the least biased
choice and
$-k{\mbox{{\rm Tr}}} \, {\hat w}_{\zeta(t)}
\log {\hat w}_{\zeta(t)} = S(\zeta(t))
$
is taken as the entropy of the macrostate. The fact that we
intend to describe isolated  systems is taken into account
by $({\hat H}-
\langle
{\hat H}
\rangle
)^2\in {\cal M}$ and assuming that:
$
\langle
({\hat H}-
\langle
{\hat H}
\rangle
)^2
\rangle^{\frac 12}
\ll
|
\langle
{\hat H}
\rangle
|
$.
The assumption that $S(\zeta(t))$ represents the entropy of
the system is physically meaningful only if the set ${\cal
M}$ is small enough: in fact if the linear span of ${\cal M}$ should be
invariant under time evolution
$S(\zeta(t))$ would be constant, contrary, for a
non-equilibrium  system, to the second principle of
thermodynamics. Due to the fact that the relevant
observables are a relatively small subset of all
observables one can safely assume that invariance of ${\cal
M}$ under time evolution does not occur for realistic
Hamiltonians; on the other hand just by this fact, the
naive identification of the statistical operator
${\hat \varrho}_t
=
        e^{-{{
        i
        \over
         \hbar
        }}{\hat H}(t-t_0)}
        {\hat \varrho}_{t_0}
        e^{+{{
        i
        \over
         \hbar
        }}{\hat H}(t-t_0)}
$
with
${\hat w}_{\zeta(t)}$ for all $t$ becomes impossible. To the
intriguing problem of the relationship between
${\hat \varrho}_t$
and ${\hat w}_{\zeta(t)}$ is devoted the next section.

At any time $t$, at least as far as the expectations of the
relevant observables are concerned, the statistical
collection is characterized in a natural way by the
parameters $\zeta(t)$: therefore one is induced to take
$\zeta(t)$ as an objective property of each member of the
statistical collection. This is indeed the case in the
typical applications of non-equilibrium statistical
mechanics. This is most typically seen in the case of the
velocity field for a continuum, related to the
expectations of relevant observables by:
$
        {\mbox{\bf v}}({\mbox{{\bf x}}},t)=
        {
        \langle
        {\hat {\mbox{\bf p}}} ({\mbox{{\bf x}}})
        \rangle_t
        /
        \langle
        {\hat \rho} ({\mbox{{\bf x}}})
        \rangle_t
        }
$.
However ${\mbox{\bf v}}({\mbox{{\bf x}}},t)$ has also an objective
meaning for each individual system, as seen in the
phenomenological description of macroscopic systems inside
mechanics of continua. Similar considerations can be made
also for other macroscopic (classical) fields, e.g., the
temperature field.
The very fact that the expectation values of
relevant observables are linked to the {\it objective}
parameters $\zeta(t)$ implies that the perturbation of the
macrosystem produced by the measurement of relevant
observables does not affect too much the expectation values
themselves: in fact physics points out that measurements not
disturbing the parameters  $\zeta(t)$ of the system are
indeed feasible. On the contrary one cannot expect that the
same situation occurs for the probability distribution
of the relevant observables (in case of fluctuating
variables). Just this different status of average values in
comparison to probability distributions is the key point to
make objectivity of macrosystem compatible with the typical
issue arising due to quantum mechanical measurement. A check
of this statement can be done inside the theory of
continuous measurement: the perturbation of the no longer
isolated macrosystem due to the measuring apparatus can be
represented by a non-Hamiltonian  contribution ${\cal
L}_{int}$ to the generator of time evolution. Then a
continuous measurement can be described of suitable
observables already specified inside ${\cal L}_{int}$,
obtaining the whole statistics for these observables.
It is seen that expectation values depend in a
regular way on  ${\cal L}_{int}$, so that in the limit
${\cal L}_{int} \rightarrow 0$ the isolated system dynamics
is recovered. On the contrary already the second momenta of
probability distributions diverge in this
limit~\cite{continue}. We intend to treat this problem in
more general way in future papers.

\section{Preparation  procedure for a  macroscopic system}
The first problem which is now to be faced in order to
construct the dynamics of a macrosystem, isolated for times
$t > t_0$, is to give its statistical operator
${\hat \varrho}_t$ and to elucidate  the relationship
between ${\hat \varrho}_t$ and the representative
${\hat w}_{\zeta(t)}$ of the macrostate $M_t$. The basic
ingredient to start with is the unitary evolution
$
        {\hat \varrho}_t
        =
        e^{-{{
        i
        \over
         \hbar
        }}{\hat H}(t-t_0)}
        {\hat \varrho}_{t_0}
        e^{+{{
        i
        \over
         \hbar
        }}{\hat H}(t-t_0)},
$
${\hat H}={\hat H}^{\scriptscriptstyle\dagger}$
being the Hamiltonian for the isolated system.
If $t_0$ is taken as {\it initial} time the most
straightforward approach would be to take
        \begin{equation}
        \label{2.2}
        {\hat \varrho}_{t_0}  =
        {\hat w}_{\zeta(t_0)};
        \end{equation}
however such an assumption is not satisfactory since the
initial time $t_0$ would have a privileged role, being in
general ${\hat \varrho}_t \not= {\hat w}_{\zeta(t)}$. Let us
stress that this is not appreciated as a problem, typically
inside {\it information thermodynamics}, if the statistical
operator is considered as representative of the {\it
information} about the  system: then at time $t_0$ the only
available information is just $M_{t_0}$, while at time $t$
information increases since $M_{t'}$, $t'\in [t_0,t]$, is in
principle known. Our standpoint about ${\hat \varrho}_t$ is
different: it represents the concrete  preparation
procedure of the  statistical collection for all times $t'\leq
t$. Then choice (\ref{2.2}) should be motivated on the basis
of the way the system was prepared at times $t' < t_0$. The
meaning of choice (\ref{2.2}) is that all the history $
\left \{
M_{t'}, t'< t_0
\right \}
$ is irrelevant for the subsequent dynamical evolution: the
fact that ${\hat \varrho}_t \not= {\hat w}_{\zeta(t)}$ for
$t>t_0$
indicates in principle that this is no longer true for $t>
t_0$; then  (\ref{2.2}) becomes the key problem. One could
check condition (\ref{2.2}) measuring the expectation values
$
\langle
{\dot {{\hat A}_j}}({\mbox{{\boldmath$\xi$}}})
\rangle
$
of the variables
$
{\dot {{\hat A}_j}}({\mbox{{\boldmath$\xi$}}})
=
{
i
\over
 \hbar
}
[
{\hat H},{\hat A}_j({\mbox{{\boldmath$\xi$}}})
]
$
and comparing with
$
{\mbox{{\rm Tr}}} \,
(
{\dot {{\hat A}_j}}({\mbox{{\boldmath$\xi$}}})
{\hat w}_{\zeta(t_0)}
)
$.
Physics indicates a profound difference between
${\hat \varrho}_t$ and ${\hat w}_{\zeta(t)}$: as far 
as the von Neumann entropy is concerned
$
-k
{\mbox{{\rm Tr}}} \,
{\hat \varrho}_t   \log
{\hat \varrho}_t
=
-k
{\mbox{{\rm Tr}}} \,
{\hat \varrho}_{t_0}   \log
{\hat \varrho}_{t_0}
$,
while 
$
-k
{\mbox{{\rm Tr}}} \,
{\hat w}_{\zeta(t)}   \log
{\hat w}_{\zeta(t)}
$
in general does increase.
As a general assumption (\ref{2.2}) would be acceptable if
$t_0$ should be a very special time ({\it big bang} time!),
when no previous history exists, or has been erased.
Actually an experimental  preparation implies a {\it
separation} of a physical system from the environment: one
must start with isolated systems (open macroscopic systems
remain of course an open physical problem) and this
preparation  is operatively associated to a finite time
interval: $[T,{t_0}]$. Then ${\hat \varrho}_{t_0}$
represents what has been done with the system in a
laboratory during the time interval $[T,{t_0}]$. Let us
observe that ${\hat \varrho}_t$, $t> t_0$ still represents a
family of preparations arising for the isolated system, due
to spontaneous time evolution: our goal is to give a
prescription to build the set ${\hat \varrho}_t$,
$t> t_0$, compatible with respect to 
the unitary evolution.
We formalize the preparation  procedure $[T,t_0]$ by sharp
measurements of $M_t$ at time points $T$ and $t_0$,
isolation of the system being achieved at time $t_0$, and by
control measurements of variables
$
\int_T^{t_0} dt' \,
{\hat A}_j({\mbox{{\boldmath$\xi$}}},t') h_\alpha (t')
$,
$h_\alpha (t')$
being suitable test functions (e.g., $h_\alpha (t)=\cos
\omega_\alpha t$). 
One expects that not only the densities
${\hat A}_j({\mbox{{\boldmath$\xi$}}},t)$ should be controlled but
also the corresponding currents
${\hat {\bf J}}_j({\mbox{{\boldmath$\xi$}}})$. To express the fact
that the system is only biased by these measurements and
controls, we use the principle of maximal entropy to
determine ${\hat \varrho}_{t_0}$. Then ${\hat \varrho}_{t_0}$
has the following structure:
        \begin{eqnarray}
        \label{2.3}
        {\hat \varrho}_{t_0}
        \!\!
        \!\!
        &=&
        \!\!
        \!\!
        \exp
        \left \{
        -\sum_j
        \int d {\mbox{{\boldmath$\xi$}}} \,
        \gamma_j ({\mbox{{\boldmath$\xi$}}},{t_0})
        {\hat A}_j({\mbox{{\boldmath$\xi$}}})
        +
        \sum_{j \alpha}
        \int d {\mbox{{\boldmath$\xi$}}}  \,
        \gamma_{j\alpha} ({\mbox{{\boldmath$\xi$}}})
        \int_T^{t_0}  dt' \,
        {\hat A}_j({\mbox{{\boldmath$\xi$}}},-(t_0 - t'))
        h_{j\alpha}(t')
        \right.
        \nonumber
        \\
        &&
        \!\!
        \!\!
        \hphantom{
        \exp
        \left \{
        \right.
        }
        +
        \sum_{j \alpha}
        \int d {\mbox{{\boldmath$\xi$}}} \,
        {\mbox{{\boldmath $\gamma$}}}_{j\alpha} ({\mbox{{\boldmath$\xi$}}})
        \cdot
        \int_T^{t_0} dt' \,
        {\hat {\bf J}}_j({\mbox{{\boldmath$\xi$}}},-(t_0 - t'))
        h_{j\alpha}(t')
        \nonumber
        \\
        &&
        \!\!
        \!\!
        \hphantom{
        \exp
        \left \{
        \right.
        }
        \left.
        -\sum_j
        \int d {\mbox{{\boldmath$\xi$}}} \,
        \gamma_j ({\mbox{{\boldmath$\xi$}}},T)
        {\hat A}_j({\mbox{{\boldmath$\xi$}}},-(t_0 -T))
        \right \}
        .
        \end{eqnarray}
Let us express the decisive role of the measurement of the 
relevant observables ${\hat A}_j({\mbox{{\boldmath$\xi$}}})$ at
time $t_0$, assuming that
        \begin{equation}
        \label{2.4}
        \gamma_j ({\mbox{{\boldmath$\xi$}}},{t_0})
        =
        \zeta_j ({\mbox{{\boldmath$\xi$}}},{t_0})
        \end{equation}
where $
\left \{
\zeta(t_0)
\right \}
$
is the macroscopic state of the system. We call a
preparation  procedure a {\it suitable  preparation procedure} if
(\ref{2.4}) is satisfied. Let us observe that due to the
asymmetry between
$
        \gamma_j ({\mbox{{\boldmath$\xi$}}},{t})
$
and
$
        \gamma_j ({\mbox{{\boldmath$\xi$}}},{T})
$
introduced by (\ref{2.4}) a time arrow is introduced in a
very clear way.
Among the constants of motion ${\hat C}_l$ a particular role has
the identity operator; its  contribution to the exponential
(\ref{2.3}) accounts for normalization as it was indicated
in \S~1.
The time evolution of ${\hat \varrho}_{t_0}$ is
straightforward:
        \begin{eqnarray}
        \label{2.6}
        {\hat \varrho}_t
        =
        e^{-{{
        i
        \over
         \hbar
        }}{\hat H}(t-t_0)}
        {\hat \varrho}_{t_0}
        e^{+{{
        i
        \over
         \hbar
        }}{\hat H}(t-t_0)}
        \!\! \!\! 
        &=& 
        \!\! \!\! 
        \exp
        \left \{
        {}  
        -\zeta_0 (t) {\hat {\bf 1}}
        -\sum_j
        \int d {\mbox{{\boldmath$\xi$}}} \,
        \zeta_j ({\mbox{{\boldmath$\xi$}}},{t_0})
        {\hat A}_j({\mbox{{\boldmath$\xi$}}},-(t-t_0))
        \right.
        \nonumber
        \\
        &&
        \hphantom{ \!\! \!
        \exp
        \left \{
        \right.
        }
        +
        \sum_{j \alpha}
        \int d {\mbox{{\boldmath$\xi$}}} \,
        \gamma_{j\alpha} ({\mbox{{\boldmath$\xi$}}})
        \int_T^{t_0} dt'\,
        {\hat A}_j({\mbox{{\boldmath$\xi$}}},-(t- t'))
        h_{j\alpha}(t')
        \nonumber
        \\
        &&
        \hphantom{ \!\! \! 
        \exp
        \left \{
        \right.
        }
        +
        \sum_{j \alpha}
        \int d {\mbox{{\boldmath$\xi$}}} \,
        {\mbox{{\boldmath $\gamma$}}}_{j\alpha} ({\mbox{{\boldmath$\xi$}}})
        \cdot
        \int_T^{t_0} dt'\,
        {\hat {\bf J}}_j({\mbox{{\boldmath$\xi$}}},-(t - t'))
        h_{j\alpha}(t')
        \nonumber
        \\
        &&
        \hphantom{ \!\! \!
        \exp
        \left \{
        \right.
        }
        \left.
        -
        \sum_j
        \int d {\mbox{{\boldmath$\xi$}}} \,
        \gamma_j ({\mbox{{\boldmath$\xi$}}},T)
        {\hat A}_j({\mbox{{\boldmath$\xi$}}},-(t -T))
        \right \}
        ,
        \end{eqnarray}
then one can determine the macrostate at time $t$, using the
expectations
$C_{lt}=C_{l t_0}$,
$
\langle
{\hat A}_j({\mbox{{\boldmath$\xi$}}})
\rangle_t
=
{\mbox{{\rm Tr}}} \,
(
{\hat A}_j({\mbox{{\boldmath$\xi$}}})
{\hat \varrho}_t
)
$,
and solving the equations
        \[
        {\mbox{{\rm Tr}}} \,
        \left(
        {\hat C}_l
        {\hat w}_{\zeta{(t)}}
        \right)
        =
        C_{lt_0}
        , \qquad
        {\mbox{{\rm Tr}}} \,
        \left(
        {\hat A}_j({\mbox{{\boldmath$\xi$}}})
        {{\hat w}_{\zeta{(t)}}}
        \right)
        =
        \langle
        {\hat A}_j({\mbox{{\boldmath$\xi$}}})
        \rangle_t
        ,
        \]
according to the definition given in \S~1.
Let us now rewrite
$
        \zeta_j({\mbox{{\boldmath$\xi$}}},t_0)
        {\hat A}_j({\mbox{{\boldmath$\xi$}}},-(t-t_0))
$ using the $\zeta_j({\mbox{{\boldmath$\xi$}}},t)$
determined in this way:
        \begin{eqnarray}
        \label{2.8}
        &&
        \zeta_j({\mbox{{\boldmath$\xi$}}},t_0)
        {\hat A}_j({\mbox{{\boldmath$\xi$}}},-(t-t_0))
        =
        \zeta_j({\mbox{{\boldmath$\xi$}}},t)
        {\hat A}_j({\mbox{{\boldmath$\xi$}}})
        -
        \int_{t_0}^t
        dt' \,
        {
        d
        \over
         dt'
        }
        \left[
        \zeta_j({\mbox{{\boldmath$\xi$}}},t')
        {\hat A}_j({\mbox{{\boldmath$\xi$}}},-(t-t'))
        \right]
        \\
        &&
        =
        \zeta_j({\mbox{{\boldmath$\xi$}}},t)
        {\hat A}_j({\mbox{{\boldmath$\xi$}}})
        -
        \int_{t_0}^t
        dt' \,
        {\dot{\zeta}}_j({\mbox{{\boldmath$\xi$}}},t')
        {\hat A}_j({\mbox{{\boldmath$\xi$}}},-(t-t'))
        -
        \int_{t_0}^t
        dt' \,
        \zeta_j({\mbox{{\boldmath$\xi$}}},t')
        {\dot {{\hat A}_j} }({\mbox{{\boldmath$\xi$}}},-(t-t'))
        \nonumber
        \\
        &&
        =
        \zeta_j({\mbox{{\boldmath$\xi$}}},t)
        {\hat A}_j({\mbox{{\boldmath$\xi$}}})
        -
        \int_{t_0}^t
        dt' \,
        {\dot{\zeta}}_j({\mbox{{\boldmath$\xi$}}},t')
        {\hat A}_j({\mbox{{\boldmath$\xi$}}},-(t-t'))
        +
        \int_{t_0}^t
        dt' \,
        \zeta_j({\mbox{{\boldmath$\xi$}}},t')
        div {\hat {\bf J}}_j({\mbox{{\boldmath$\xi$}}},-(t-t'))
        .
        \nonumber
        \end{eqnarray}
Replacing  (\ref{2.8}) inside (\ref{2.6}) one has:
        \begin{eqnarray}
        \label{2.9}
        {\hat \varrho}_t
        \!\!
        &=&
        \!\!
        \exp
        \left \{
        {}
        -\zeta_0(t)  {\hat {\bf 1}}
        -\sum_j
        \int d {\mbox{{\boldmath$\xi$}}} \,
        \zeta_j ({\mbox{{\boldmath$\xi$}}},{t})
        {\hat A}_j({\mbox{{\boldmath$\xi$}}})
        +
        \sum_j
        \int_{t_0}^t dt'
        \int d {\mbox{{\boldmath$\xi$}}} \,
        {\dot{\zeta}}_j ({\mbox{{\boldmath$\xi$}}},{t'})
        {\hat A}_j({\mbox{{\boldmath$\xi$}}},-(t- t'))
        \right.
        \nonumber
        \\
        \!\!
        &&
        \hphantom{
        \exp
        \left \{
        \right.
        }
        {}
        +
        \sum_{j \alpha}
        \int^{t_0}_T dt'
        \int d {\mbox{{\boldmath$\xi$}}} \,
        \gamma_{j\alpha} ({\mbox{{\boldmath$\xi$}}})
        {\hat A}_j({\mbox{{\boldmath$\xi$}}},-(t- t'))
        h_{j\alpha}(t')
        \nonumber
        \\
        \!\!
        &&
        \hphantom{
        \exp
        \left \{
        \right.
        }
        {}
        -
        \sum_j
        \int_{t_0}^t dt'
        \int d {\mbox{{\boldmath$\xi$}}} \,
        {{\zeta}}_j ({\mbox{{\boldmath$\xi$}}},{t'})
        {div {\hat {\bf J}}}_j({\mbox{{\boldmath$\xi$}}},-(t- t'))
        \nonumber
        \\
        \!\!
        &&
        \hphantom{
        \exp
        \left \{
        \right.
        }
        {}
        +
        \sum_{j \alpha}
        \int_T^{t_0} dt'
        \int d {\mbox{{\boldmath$\xi$}}} \,
        {\mbox{{\boldmath $\gamma$}}}_{j\alpha} ({\mbox{{\boldmath$\xi$}}})
        \cdot
        {\hat {\bf J}}_j({\mbox{{\boldmath$\xi$}}},-(t - t'))
        h_{j\alpha}(t')
        \nonumber
        \\
        \!\!
        &&
        \hphantom{
        \exp
        \left \{
        \right.
        }
        {}
        \left.
        -
        \sum_j
        \int d {\mbox{{\boldmath$\xi$}}} \,
        \gamma_j ({\mbox{{\boldmath$\xi$}}},T)
        {\hat A}_j({\mbox{{\boldmath$\xi$}}},-(t -T))
        \right \}
        .
        \end{eqnarray}
Comparing ${\hat \varrho}_t$ given by (\ref{2.9}) with
${\hat \varrho}_{t_0}$ given by (\ref{2.3}) one observes
that the basic structure is preserved:
${\hat \varrho}_t$ represents a  preparation  procedure
terminating at time $t$, which replaces $t_0$, the initial
macrostate parameters
$\zeta_j({\mbox{{\boldmath$\xi$}}},t_0)$ being
replaced by $\zeta_j({\mbox{{\boldmath$\xi$}}},t)$. The
contribution representing the past history now extends from
$T$ to $t$ and a new part is displayed, related to the time
interval $[t_0,t]$. In place of the parameters
$
\sum_{\alpha}
\gamma_{j\alpha}({\mbox{{\boldmath$\xi$}}}) h_{j\alpha}(t')
$
which described the preparation  procedure in the time
interval $[T,t_0]$, now the parameters ${\dot
{\zeta}}_j({\mbox{{\boldmath$\xi$}}},t)$ appear, in place of the
term
$
        \sum_{j \alpha}
        {\mbox{{\boldmath $\gamma$}}}_{j\alpha} ({\mbox{{\boldmath$\xi$}}})
        \cdot
        {\hat {\bf J}}_j({\mbox{{\boldmath$\xi$}}},-(t - t'))
        h_{j\alpha}(t')
$
one deals with
$
        -\sum_j
        {{\zeta}}_j ({\mbox{{\boldmath$\xi$}}},{t'})
        {div {\hat {\bf J}}}_j({\mbox{{\boldmath$\xi$}}},-(t- t'))
$.
In this way a solution is given to the problem of justifying
assumption ${\hat \varrho}_{t_0}$ for the description of a
preparation  procedure of a macrosystem. Furthermore the
structure (\ref{2.6}) of ${\hat \varrho}_t$ also suggests a
practical solution method, that will be discussed in \S~3.
Not only ${\hat \varrho}_t$ provides by construction the
expectation values of the relevant observables through the
expressions
$
        {\mbox{{\rm Tr}}} \,
        (
        {\hat A}_j({\mbox{{\boldmath$\xi$}}}) {\hat \varrho}_t
        )
$
which are linked to the objective state parameter, but also
provides probability distributions for measurements which
can be performed on the system by the usual tools of
quantum mechanics (in the most refined case instruments or
operation valued measures).
The result (\ref{2.6}) is very close to the {\it
non-equilibrium statistical operator} proposed by
Zubarev~\cite{Zubarev}; the main difference is the choice
$T\rightarrow -\infty$ in the formulation of Zubarev, linked
to the fact that no clear distinction is introduced between
preparation  of the system and its spontaneous evolution.
The limit $T\rightarrow -\infty$ presupposes a thermodynamic
limit and introduces big difficulties for a non-equilibrium
system.

\section{Evolution equation for the classical state
parameters
\boldmath
$\{\zeta(t)\}$
\unboldmath}
The structure (\ref{2.9}) strongly suggests that the first
terms in the argument of the exponential are more
important than the remaining ones, since they alone already
determine the expectations of the relevant observables.
Then a perturbation theory becomes very natural in which the
last part of the exponential is treated as a perturbation,
the typical {\it cumulant
expansion}. The first  contributions are:
        \begin{eqnarray}
        \label{3.1}
        {
        {\mbox{{\rm Tr}}} \,
        {\hat C}
        e^{{\hat A}+{\hat B}}
        \over
        {\mbox{{\rm Tr}}} \,
        e^{{\hat A}+{\hat B}}
        }
        &=&
        {
        {\mbox{{\rm Tr}}} \,
        {\hat C}
        e^{{\hat A}}
        \over
        {\mbox{{\rm Tr}}} \,
        e^{{\hat A}}
        }
        +
        {
        {\mbox{{\rm Tr}}} \,
        {\hat C}
        \int_0^1 du \,
        e^{u{\hat A}}
        {\hat B}
        e^{(1-u){\hat A}}
        \over
        {\mbox{{\rm Tr}}} \,
        e^{{\hat A}}
        }
        -
        {
        {\mbox{{\rm Tr}}} \,
        {\hat C}
        e^{{\hat A}}
        \over
        {\mbox{{\rm Tr}}} \,
        e^{{\hat A}}
        }
        {
        {\mbox{{\rm Tr}}} \,
        {\hat B}
        e^{{\hat A}}
        \over
        {\mbox{{\rm Tr}}} \,
        e^{{\hat A}}
        }
        +
        \ldots
        .
        \end{eqnarray}
The typical equation yielding the time evolution of the
macrostate is given in terms of ${\hat \varrho}_t[\zeta]$
by the condition, arising from the definition of macrostate:
        \begin{equation}
        \label{3.2}
        {
        d
        \over
         dt
        }
        {\mbox{{\rm Tr}}} \,
        (
        {\hat A}_j({\mbox{{\boldmath$\xi$}}})
        {\hat \varrho}_t[\zeta]
        )
        =
        {
        d
        \over
         dt
        }
        {\mbox{{\rm Tr}}} \,
        (
        {\hat A}_j({\mbox{{\boldmath$\xi$}}})
        {{\hat w}_{\zeta{(t)}}}
        )
        =
        -
        \sum_{l}
        \int d {\mbox{{\boldmath$\xi$}}}' \,
        \mbox{\boldmath $\langle$}
        {\hat A}_j({\mbox{{\boldmath$\xi$}}})   ,
        {\hat A}_{l}({\mbox{{\boldmath$\xi$}}}')
        \mbox{\boldmath $\rangle$}_{{{\hat w}_{\zeta{(t)}}}}
        {\dot {\zeta}}_{l}({\mbox{{\boldmath$\xi$}}}',t),
        \end{equation}
where the Kubo correlation function has been introduced
        \[
        \mbox{\boldmath $\langle$}
        {\hat A},{\hat B}
        \mbox{\boldmath $\rangle$}_{{{\hat w}_{\zeta{(t)}}}}
        \equiv
        {\mbox{{\rm Tr}}} \,
        {\hat A}
        \int_0^1 du \,
        e^{
        u
        {\hat C}(t)
        }
        {\hat B}
        e^{
        (1-u)
        {\hat C}(t)
        }
        -
        {
        {\mbox{{\rm Tr}}} \,
        {\hat A}{{\hat w}_{\zeta{(t)}}}
        }
        {
        {\mbox{{\rm Tr}}} \,
        {\hat B}{{\hat w}_{\zeta{(t)}}}
        }
        ,
        \]
with
$
{\hat C}(t)=
        \left[
        {
        {
        -\zeta_0(t) {\hat {\bf 1}}
        -{
        \sum_j
        \int d {\mbox{{\boldmath$\xi$}}} \,
        {\zeta}_j({\mbox{{\boldmath$\xi$}}},t)
        {\hat A}_j({\mbox{{\boldmath$\xi$}}})
        }}
        }
        \right]
$.
Since
$
d 
{\hat \varrho}_t
/ dt
=
-
{
i
\over
 \hbar
}
[
{\hat H},{\hat \varrho}_t
]
$
the first term of (\ref{3.2}) becomes
$
        {\mbox{{\rm Tr}}} \,
(
{
i
\over
 \hbar
}
[
{\hat H},
{\hat A}_j({\mbox{{\boldmath$\xi$}}})
]
{\hat \varrho}_t [\zeta]
)
$. Let us represent ${\hat \varrho}_t$ in the form
(\ref{3.1}), then the following evolution equation for the
state parameters arises:
        \begin{eqnarray}
        \label{3.3}
        \lefteqn{
        -  
        \sum_{l}
        \int d {\mbox{{\boldmath$\xi$}}}' \,
        \mbox{\boldmath $\langle$}
        {\hat A}_j({\mbox{{\boldmath$\xi$}}})   ,
        {\hat A}_{l}({\mbox{{\boldmath$\xi$}}}')
        \mbox{\boldmath $\rangle$}_{{{\hat w}_{\zeta{(t)}}}}
        {\dot {\zeta}}_{l}({\mbox{{\boldmath$\xi$}}}',t)
        =
        }
        \\
        &&
        {\mbox{{\rm Tr}}} \,
        \left(
        {
        i
        \over
         \hbar
        }
        [
        {\hat H},
        {\hat A}_j({\mbox{{\boldmath$\xi$}}})
        ]
        {{\hat w}_{\zeta{(t)}}}
        \right)
        +
        \int_T^t dt'\,
        \mbox{\boldmath $\langle$}
        {
        i
        \over
         \hbar
        }
        [
        {\hat H},
        {\hat A}_j({\mbox{{\boldmath$\xi$}}})
        ]
        ,
        {\hat {\cal S}}(t')
        \mbox{\boldmath $\rangle$}_{{{\hat w}_{\zeta{(t)}}}}
        \nonumber
        \\
        &&
        -
        \sum_{l}
        \int d {\mbox{{\boldmath$\xi$}}}' \,
        \mbox{\boldmath $\langle$}
        {
        i
        \over
         \hbar
        }
        [
        {\hat H},
        {\hat A}_j({\mbox{{\boldmath$\xi$}}})
        ]
        ,
        {\hat A}_{l}({\mbox{{\boldmath$\xi$}}}',-(t-T))
        \mbox{\boldmath $\rangle$}_{{{\hat w}_{\zeta{(t)}}}}
        { {\gamma}}_{l}({\mbox{{\boldmath$\xi$}}}',T)
        + \ldots
        \nonumber
        \end{eqnarray}
In these equations the whole history of the system arises
represented by the term
$
        {\hat {\cal S}}(t')
$
and by
$
        { {\gamma}}_{l}({\mbox{{\boldmath$\xi$}}}',T)
$, where
$
        {\hat {\cal S}}(t')
$
is given by
        \begin{equation}
        \label{3.4}
        \left \{  
        \!\!
        \begin{array}{ll}  
        \sum_{j \alpha}
        \int d {\mbox{{\boldmath$\xi$}}}' \,
        [
        \gamma_{j\alpha} ({\mbox{{\boldmath$\xi$}}}')
        {\hat A}_j({\mbox{{\boldmath$\xi$}}}',-(t- t'))
        +
        {\mbox{{\boldmath $\gamma$}}}_{j\alpha}
        ({\mbox{{\boldmath$\xi$}}}')
        \cdot
        {\hat {\bf J}}_j({\mbox{{\boldmath$\xi$}}}',-(t - t'))
        ]
        h_{j\alpha}(t')
        & T\leq t'\leq t_0
        \\  
        \sum_j
        \int d {\mbox{{\boldmath$\xi$}}}'\,
        [
        {\dot{\zeta}}_j ({\mbox{{\boldmath$\xi$}}}',{t'})
        {\hat A}_j({\mbox{{\boldmath$\xi$}}}',-(t- t'))
        -
        {{\zeta}}_j ({\mbox{{\boldmath$\xi$}}}',{t'})
        {div {\hat {\bf J}}}_j({\mbox{{\boldmath$\xi$}}}',-(t- t'))        
        ]
        & t_0\leq t'\leq t
        \end{array}  
        \right.
        .
        \end{equation}
It is seen that time evolution provides an additional
preparation represented by the parameters
${\dot {\zeta}}_j({\mbox{{\boldmath$\xi$}}}',t')$ and
$\zeta_j({\mbox{{\boldmath$\xi$}}}',t')$;
thus (\ref{3.3}) are integrodifferential evolution equations
for
$\zeta_j({\mbox{{\boldmath$\xi$}}},t)$. Formally the whole memory
of the macrostate for $T\leq t'\leq t$ appears inside the
expression of
${\dot {\zeta}}_j({\mbox{{\boldmath$\xi$}}}',t')$ through the
correlation functions:
        \[
        \mbox{\boldmath $\langle$}
        {
        i
        \over
         \hbar
        }
        [
        {\hat H},
        {\hat A}_j({\mbox{{\boldmath$\xi$}}})
        ]
        ,
        {\hat A}_{l}({\mbox{{\boldmath$\xi$}}}',-(t-t'))
        \mbox{\boldmath $\rangle$}_{{{\hat w}_{\zeta{(t)}}}}
        ,
        \quad
        \mbox{\boldmath $\langle$}
        {
        i
        \over
         \hbar
        }
        [
        {\hat H},
        {\hat A}_j({\mbox{{\boldmath$\xi$}}})
        ]
        ,
        {\hat {\bf J}}_l({\mbox{{\boldmath$\xi$}}}',-(t-t'))
        \mbox{\boldmath $\rangle$}_{{{\hat w}_{\zeta{(t)}}}}
        , \
        T \leq t'\leq t.
        \]
By the first order approximation inside (\ref{3.1}) the
whole dynamics of the macrostate is controlled through {\it
two point} Kubo correlation functions  for relevant
observable and their time derivatives (currents).
Higher order correlation functions are introduced in the
higher approximations. Actually most applications of
non-equilibrium thermodynamics already fit inside the
first order approximation scheme.
Now one can take advantage from a general feature of
correlation functions: they have a {\it decaying} behavior
in time. The basic assumption about the preparation
procedure, that was assumed to be restricted inside the
finite time interval $[T,t_0]$, could be consistently assumed
also at later times. Indicating qualitatively $\tau$ as
typical decay time of the correlation functions, the
structure of ${\hat \varrho}_t$ appears justified if $t_0-T
\geq \tau$; furthermore one expects that the r.h.s. of
(\ref{3.3}) can be simplified for $t-t_0 >\tau$, dropping
the last term and replacing the integration interval $[T,t]$
with $[t-\tau,t]$, in this way providing a universal
character to the time evolution equations by elimination of
the preparation parameters.
Decaying behavior
of correlation functions is a central
issue in statistical mechanics often  achieved making use
of the thermodynamic limit, while in our case a less
schematic and more sophisticated attitude must be taken;
in fact we are considering a confined system, separated and
isolated from the environment: then the Hamiltonian has a
point spectrum and correlation functions have a
quasiperiodical time behavior. Loss of memory arises
through an interplay between choice of observables and
characterization of suitable preparation  procedures. Let us
observe that correlation functions always appear inside
time integrals: by these integrations the quasiperiodical
behavior of correlation functions can very well produce a
decaying behavior, provided other factors inside the time
integrals are smooth enough. Looking at (\ref{3.3}) and
(\ref{3.4}) these factors are seen to be the functions
$h_{j\alpha}(t')$ and the state parameters
$\zeta_j({\mbox{{\boldmath$\xi$}}},t')$, which are linked to the
expectation values $
\langle
{\hat A}_j({\mbox{{\boldmath$\xi$}}})
\rangle
$.
Also the integration on ${\mbox{{\boldmath$\xi$}}}$ will have a
smoothing effect, provided the state parameters are
homogeneous enough, which in turn depends on suitable
choice of space-time variation scale of the relevant
observables.
Let us note that neglecting in (\ref{3.3}) history in the
time interval $[T,t-\tau]$, $\tau$ being the previously
introduced typical decay time, amounts to take at time
$t-\tau$ an initial state ${\hat \varrho}_{t-\tau}={\hat
w}_{\zeta{(t-\tau)}}$, so that
${\hat \varrho}_t
=
        e^{-{{
        i
        \over
         \hbar
        }}{\hat H}\tau}
        {\hat w}_{\zeta{(t-\tau)}}
        e^{+{{
        i
        \over
         \hbar
        }}{\hat H}\tau}
$, leading in a straightforward way to the iterated
inequality:
$
        S_t
        \geq S_{t-\tau}
        \geq S_{t-2\tau}
        \geq \ldots S_{t- r\tau}  
$,
loosely a {\it mean stepwise increase of entropy}. As long
as ${\hat \varrho}_t$ is not a Gibbs state one has the
strict inequality $S_t > S_{t-\tau}$: this is indeed the
case, due to the history  contribution. Since by the
constants of motion
$
A_j =
        \int d {\mbox{{\boldmath$\xi$}}}
        \langle
        {\hat A}_j({\mbox{{\boldmath$\xi$}}})
        \rangle_t
$
the finite bound
$
S
(
\{
{\hat A}_j({\mbox{{\boldmath$\xi$}}})=
{
{\hat A}_j
/
          V
}
 \}
)
\geq
S
(
 \{
{\hat A}_j({\mbox{{\boldmath$\xi$}}})
 \}
)
$
is put on the entropy, one can in this way conclude that the
system approaches for $t\rightarrow +\infty$ the Gibbs state,
determined by the constants of motion: i.e., its equilibrium
state.
To take further into account the particular role of the
constants of motion, let us replace the relevant density
field ${\hat A}_j({\mbox{{\boldmath$\xi$}}})$ with a suitable set
of transformed variables
        \[
        {{\hat a}}_{j  n} =
        \int d {\mbox{{\boldmath$\xi$}}} \,
        u^*_{n}({\mbox{{\boldmath$\xi$}}})
        {\hat A}_j({\mbox{{\boldmath$\xi$}}})    ,
        \]
where $u_{n}({\mbox{{\boldmath$\xi$}}})$ is some suitable
complete orthonormal set of functions (e.g., Fourier
functions) defined in the phase space region on which the
fields are defined.
Let us assume that $u_0 ({\mbox{{\boldmath$\xi$}}})$ is constant so
that the variables ${\hat a}_{j 0}$ we have related to
densities of conserved quantities, coincide with the
conserved observables; then (\ref{3.3}) becomes:
        \begin{eqnarray}
        \label{3.7}
        -
        \sum_{j'n'}
        \mbox{\boldmath $\langle$}
        {\hat a}_{j0},
        {\hat a}_{j'n'}
        \mbox{\boldmath $\rangle$}_{{{\hat w}_{\zeta{(t)}}}}
        {\dot {\zeta}}_{j'n'}(t)
        \!\!
        &=&
        \!\!
        0
        \\
        -
        \sum_{j'n'}
        \mbox{\boldmath $\langle$}     
        {\hat a}_{j n},
        {\hat a}_{j'n'}
        \mbox{\boldmath $\rangle$}_{{{\hat w}_{\zeta{(t)}}}}
        {\dot {\zeta}}_{j'n'}(t)
        \!\!
        &=&
        \!\!
        {\mbox{{\rm Tr}}} \,
        \left(
        {
        i
        \over
         \hbar
        }
        [
        {\hat H},
        {\hat a}_{jn}
        ]
        {{\hat w}_{\zeta{(t)}}}
        \right)
        +
        \int_T^t
        dt'\,
        \mbox{\boldmath $\langle$}
        {
        i
        \over
         \hbar
        }
        [
        {\hat H},
        {\hat a}_{jn}
        ]
        ,
        {\hat {\cal S}}(t')
        \mbox{\boldmath $\rangle$}_{{{\hat w}_{\zeta{(t)}}}}
        \nonumber
        \\
        &&
        \!\!
        -
        \sum_{j'n'}
        \mbox{\boldmath $\langle$}
        {
        i
        \over
         \hbar
        }
        [
        {\hat H},
        {\hat a}_{jn}
        ]
        ,
        {\hat a}_{j'n'}(-(t-T))
        \mbox{\boldmath $\rangle$}_{{{\hat w}_{\zeta{(t)}}}}
        { {\gamma}}_{j'n'}(T),
        \  n>0
        \nonumber
        \end{eqnarray}
where
$
        {\hat {\cal S}}(t')
$
can be easily rewritten in terms of the new variables.
The general problem of extracting a system of
integrodifferential equations for $\zeta_{jn}(t)$ can be
solved restricting to a finite subset of variables
$\{ {\hat a}_{jn} \}_{n\leq N}$, and taking the inverse of the
matrix
$
\mbox{\boldmath $\langle$}
{\hat a}_{jn},
{\hat a}_{j'n '}
\mbox{\boldmath $\rangle$}_{{{\hat w}_{\zeta{(t)}}}}
$.

\section{Dynamical semigroup description}
A separate role is now given to the constants of motion
${\hat a}_{j0}$, having constant expectations and in this
way influencing the state variables through the first line of 
(\ref{3.7}), as
compared to the other observables, which drive the dynamics
through the second line of 
(\ref{3.7}). It will turn out to be useful to
formalize in the following way these different roles: by
means of the Gibbs state ${\hat w}_{\zeta(t)}$ the following
sesquilinear form on the space of operators~\cite{scalare}
can be defined
by
$
        \mbox{\boldmath $\langle$}
        {\hat A},{\hat B}
        \mbox{\boldmath $\rangle$}_{{{\hat w}_{\zeta{(t)}}}}
$, by which a time-dependent Hilbert space structure on the
space of operators can be introduced, linked to the
macrostate at time $t$. Let us consider the subspace spanned
by the constants of motion ${\hat a}_{j0}$ and decompose the
observables
${\hat a}_{jn}$, $n\geq 1$ in a parallel and orthogonal
component with respect to this subspace. Obviously we can
restrict the study of time evolution to these orthogonal
components
${\hat a}_{jn{\scriptscriptstyle\perp}}$, $n\geq 1$.
One can expect that if the state parameters solely depend on
time, during the relevant part of history, differential
equations for time evolution instead of the
integrodifferential equations discussed in \S~3 should
arise. Let us treat (\ref{2.9}) rewritten in terms of the
variables ${\hat a}_{jn}$ in a slightly different way:
        \begin{eqnarray*}
        {\hat \varrho}_t
        \!\!\!
        &=&
        \!\!\!
        \exp
        \left \{
        -\sum_{j n} \zeta_{jn}(t){\hat a}_{jn}
        +
        \sum_{j n\geq 1}
        \int_{t_0}^t dt' \,
        {\dot{\zeta}}_{jn}(t')
        {\hat a}_{jn}(-(t- t'))
        \right.
        \\
        \!\!\!
        &&
        \!\!\!
        \hphantom{
        \exp
        \left \{
        \right.
        }
        +
        \!\!
        \sum_{j\alpha  n\geq 1 }
        \int_T^{t_0} dt' \,
        \gamma_{j\alpha n}(t')
        {\hat a}_{jn} (-(t- t')) h_{j\alpha}(t')
        +
        \sum_{j n\geq 1}
        \int_{t_0}^t dt' \,
        {{\zeta}}_j ({t'})
        {\dot {{\hat a}}}_{jn} (-(t- t'))
        \nonumber
        \\
        \!\!\!
        &&
        \!\!\!
        \hphantom{
        \exp
        \left \{
        \right.
        }
        \left.
        +
        \!\!
        \sum_{j\alpha n\geq 1 }
        \int_T^{t_0} dt' \,
        {\mbox{{\boldmath $\gamma$}}}_{j\alpha n}
        \cdot
        {\hat {\bf J}}_{jn}(-(t - t'))
        h_{j\alpha}(t')
        -
        \sum_{j n\geq 1}
        \gamma_{jn} (T){\hat a}_{jn} (-(t -T))
        \right \}
        .
        \nonumber
        \end{eqnarray*}
The state parameters are given by the equations:
$
        {\mbox{{\rm Tr}}} \,
        {\hat a}_{jn} {{\hat w}_{\zeta{(t)}}}
        =
        {\mbox{{\rm Tr}}} \,
        {\hat a}_{jn} {\hat \varrho}_t 
$;
let us now look whether a differential equation for $\zeta(t)$ can
be derived from these
        \begin{eqnarray}
        \label{4.3}
        {\mbox{{\rm Tr}}} \,
        {\hat a}_{jn{\scriptscriptstyle\parallel}} {\hat \varrho}_t
        \!\!
        &=&
        \!\!
        {\mbox{{\rm Tr}}} \,
        {\hat a}_{jn{\scriptscriptstyle\parallel}} {{\hat w}_{\zeta{(t)}}}
        =
        {E}_{jn{\scriptscriptstyle\parallel}}
        \\
        {\mbox{{\rm Tr}}} \,
        {\hat a}_{jn{\scriptscriptstyle\perp}} {{\hat w}_{\zeta{(t)}}}
        \!\!
        &=&
        \!\!
        {\mbox{{\rm Tr}}} \,
        {\hat a}_{jn{\scriptscriptstyle\perp}} {{\hat w}_{\zeta{(t)}}}
        +
        \sum_{j'n'\geq1}
        \int_{t_0}^t
        dt' \,
        \left(
        \mbox{\boldmath $\langle$}
        {\hat a}_{jn{\scriptscriptstyle\perp}}
        ,
        {\hat a}_{j'n'}(-(t-t'))
        \mbox{\boldmath $\rangle$}_{{{\hat w}_{\zeta{(t)}}}}
        {\dot {\zeta}}_{j'n'}(t')
        \right.
        \nonumber
        \\
        &&
        \!\!
        \left.
        \hphantom{
        {\mbox{{\rm Tr}}} \,
        {\hat a}_{jn{\scriptscriptstyle\perp}} {{\hat w}_{\zeta{(t)}}}
        +
        \sum_{j'n'\geq1}
        \int_{t_0}^t
        dt' \,
        \left(
        \right.
        }
        +
        \mbox{\boldmath $\langle$}
        {\hat a}_{jn{\scriptscriptstyle\perp}}
        ,
        {\dot {{\hat a}}}_{j'n'}(-(t-t'))
        \mbox{\boldmath $\rangle$}_{{{\hat w}_{\zeta{(t)}}}}
        {{\zeta}}_{j'n'}(t')
        \right)
        \nonumber
        \\
        &&
        \!\!
        +
        \sum_{j'\alpha n'\geq 1}
        \int_{t_0}^t dt' \,
        \mbox{\boldmath $\langle$}
        {\hat a}_{jn{\scriptscriptstyle\perp}}
        ,
        {\hat a}_{j'n'}(-(t-t'))
        \mbox{\boldmath $\rangle$}_{{{\hat w}_{\zeta{(t)}}}}
        \gamma_{j'\alpha n'}(t')
        h_{j'\alpha}(t')
        \nonumber
        \\
        &&
        \!\!
        +
        \sum_{j'\alpha n'\geq 1}
        \int^{t_0}_T dt' \,
        \mbox{\boldmath $\langle$}
        {\hat a}_{jn{\scriptscriptstyle\perp}}
        ,
        {\hat {\bf J}}_{j'n'}(-(t - t'))
        \mbox{\boldmath $\rangle$}_{{{\hat w}_{\zeta{(t)}}}}
        \cdot
        {\mbox{{\boldmath $\gamma$}}}_{j'\alpha n'}
        h_{j'\alpha}(t')
        \nonumber
        \\
        &&
        \!\!        -
        \sum_{j' n'\geq 1}
        \mbox{\boldmath $\langle$}
        {\hat a}_{jn{\scriptscriptstyle\perp}}
        ,
        {\hat a}_{j'n'} (-(t -T))
        \mbox{\boldmath $\rangle$}_{{{\hat w}_{\zeta{(t)}}}}
        \gamma_{j'n'} (T)
        .
        \nonumber
        \end{eqnarray}
In (\ref{4.3}) we have taken into account that the
contribution of $n' =0$ can be dropped due to orthogonality
with ${\hat a}_{jn{\scriptscriptstyle\perp}}$ ($n\geq 1$).
As we already did in \S~3, let us use the time decay of the
history of the macrosystem, characterized by the typical
time $\tau$; if $t-t_0>\tau$, we can neglect the preparation
part represented by the last three terms of (\ref{4.3}) and
we can rewrite the second part of (\ref{4.3}) in the form
        \begin{equation}
        \label{4.4}
        \sum_{j'n'\geq 1} \!
        \frac 1\tau
        \int_{0}^\tau
        d\tau' \,
        [
        \mbox{\boldmath $\langle$}
        {\hat a}_{jn{\scriptscriptstyle\perp}}
        ,
        {\hat a}_{j'n'}(-\tau')
        \mbox{\boldmath $\rangle$}_{{{\hat w}_{\zeta{(t)}}}}
        {\dot {\zeta}}_{j'n'}(t-\tau')
        +
        \mbox{\boldmath $\langle$}
        {\hat a}_{jn{\scriptscriptstyle\perp}}
        ,
        {\dot {{\hat a}}}_{j'n'}(-\tau')
        \mbox{\boldmath $\rangle$}_{{{\hat w}_{\zeta{(t)}}}}
        {{\zeta}}_{j'n'}(t-\tau')
        ] = 0
        .
        \end{equation}
The main point is now to assume that the relevant
observables have a time evolution that is slow enough; more
precisely one has for the state parameters:
        \begin{equation}
        \label{4.5}
        \zeta_{j'n'}(t-\tau') \approx \zeta_{j'n'}(t);
        \qquad
        {\dot {\zeta}}_{j'n'}(t-\tau') \approx {\dot {\zeta}}_{j'n'}(t);
        \qquad \tau' \leq \tau.
        \end{equation}
A similar property is exhibited by the expectations
and we can expect it also for correlation functions
$
        \mbox{\boldmath $\langle$}
        {\hat a}_{jn{\scriptscriptstyle\perp}}
        ,
        { {{\hat a}}}_{j'n'}(-\tau')
        \mbox{\boldmath $\rangle$}_{{{\hat w}_{\zeta{(t)}}}}
$.
We represent the 
first term of (\ref{4.4}) using (\ref{4.5})
        \[
        -
        \!
        \sum_{j'n'\geq 1}
        \mbox{\boldmath $\langle$}
        {\hat a}_{jn{\scriptscriptstyle\perp}}
        ,
        {\hat a}_{j'n'}
        \mbox{\boldmath $\rangle$}_{{{\hat w}_{\zeta{(t)}}}}
        {\dot {\zeta}}_{j'n'}(t)
        -
        \!
        \sum_{j'n'\geq 1}
        \frac 1\tau
        \int_{0}^\tau
        d\tau' \,
        \mbox{\boldmath $\langle$}
        {\hat a}_{jn{\scriptscriptstyle\perp}}
        ,
        (
        {{{\hat a}}}_{j'n'}(-\tau') -
        {{{\hat a}}}_{j'n'}
        )
        \mbox{\boldmath $\rangle$}_{{{\hat w}_{\zeta{(t)}}}}
        {\dot{\zeta}}_{j'n'}(t)
        =
        \]
        \[
        {
        d
        \over
         dt
        }
        {
        {\mbox{{\rm Tr}}} \,
        {\hat a}_{jn{\scriptscriptstyle\perp}}
        \exp
        \left \{
        {  -
        \sum_{lm}
        {{\zeta}}_{lm}(t)
        { {{\hat a}}}_{lm}
        }
        \right \}
        \over
        {\mbox{{\rm Tr}}} \,
        \exp
        \left \{
        { -
        \sum_{lm}
        {{\zeta}}_{lm}(t)
        { {{\hat a}}}_{lm}
        }
        \right \}
        }
        -
        \frac 1\tau
        \int_{0}^\tau
        d\tau' \,
        \mbox{\boldmath $\langle$}
        {\hat a}_{jn{\scriptscriptstyle\perp}}
        ,
        \!
        \sum_{j'n'\geq 1}
        (
        {{{\hat a}}}_{j'n'}(-\tau') -
        {{{\hat a}}}_{j'n'}
        )
        {\dot{\zeta}}_{j'n'}(t)
        \mbox{\boldmath $\rangle$}_{{{\hat w}_{\zeta{(t)}}}}
        .
        \]
The structure of the 
second term of (\ref{4.4}) is as follows:
        \begin{equation}
        \label{4.6}
        \frac 1\tau
        \int_{0}^\tau
        \!
        d\tau' \,
        \!
        \left[
        {
        {\mbox{{\rm Tr}}} \,
        {\hat A}
        \int_0^1 du \,
        e^{-u{\hat C}_t}
        {
        i
        \over
         \hbar
        }
        [
        {\hat H},
        {\hat C}_t (-\tau')
        ]
        e^{-(1-u){\hat C}_t}
        \over
        {\mbox{{\rm Tr}}} \,
        e^{-{\hat C}_t}
        }
        -
        {
        {\mbox{{\rm Tr}}} \,
        {\hat A}
        e^{-{\hat C}_t}
        \over
        {\mbox{{\rm Tr}}} \,
        e^{-{\hat C}_t}
        }
        {
        {\mbox{{\rm Tr}}} \,
        {
        i
        \over
         \hbar
        }
        [
        {\hat H},
        {\hat C}_t (-\tau')
        ]
        e^{-{\hat C}_t}
        \over
        {\mbox{{\rm Tr}}} \,
        e^{-{\hat C}_t}
        }
        \right]
        ,
        \end{equation}
where ${\hat C}_t =
        \sum_{lm}
        {{\zeta}}_{lm}(t)
        { {{\hat a}}}_{lm}
$, ${\hat A}={\hat a}_{jn{\scriptscriptstyle\perp}}$.
We replace now inside (\ref{4.6}) expressions like
$\exp \{ -\alpha{\hat C}_t \}$ by
$\exp \{ -\alpha{\hat C}_t(-\tau) -\alpha
(
{\hat C}_t(-\tau)-{\hat C}_t
)
\}$
and expand with respect
to
$
{\hat C}_t(-\tau)-{\hat C}_t
$
by a perturbative expansion retaining the first terms, so
that (\ref{4.6}) becomes:
        \[
        -
        \frac 1\tau
        \int_{0}^\tau
        d\tau' \,
        \left[
        {
        {\mbox{{\rm Tr}}} \,
        {\hat A}
        {
        i
        \over
         \hbar
        }
        [
        {\hat H},
        e^{-{\hat C}_t (-\tau')}
        ]
        \over
        {\mbox{{\rm Tr}}} \,
        e^{-{\hat C}_t}
        }
        +
        {
        {\mbox{{\rm Tr}}} \,
        {\hat A}
        e^{-{\hat C}_t}
        \over
        {\mbox{{\rm Tr}}} \,
        e^{-{\hat C}_t}
        }
        {
        {\mbox{{\rm Tr}}} \,
        {
        i
        \over
         \hbar
        }
        [
        {\hat H},
        {\hat C}_t (-\tau')
        ]
        e^{-{\hat C}_t(-\tau)}
        \over
        {\mbox{{\rm Tr}}} \,
        e^{-{\hat C}_t(-\tau)}
        }
        \right]
        ,
        \]
due to the cyclicity of the trace operation the last term
vanishes and the first one becomes:
        \[
        \frac 1\tau
        \int_{0}^\tau
        d\tau' \,
        {
        {\mbox{{\rm Tr}}} \,
        {\dot{ {\hat A}_{\hphantom{j}} }}(\tau')
        e^{-{\hat C}_t}
        \over
        {\mbox{{\rm Tr}}} \,
        e^{-{\hat C}_t}
        }
        =
        \frac 1\tau
        {
        {\mbox{{\rm Tr}}} \,
        \left(
        {{{\hat A}}}(\tau)
        -
        {{{\hat A}}}(0)
        \right)
        e^{-{\hat C}_t}
        \over
        {\mbox{{\rm Tr}}} \,
        e^{-{\hat C}_t}
        } .
        \]
The first order result is:
        \begin{equation}
        \label{4.7}
        {
        d
        \over
         dt
        }
        {
        {\mbox{{\rm Tr}}} \,
        {\hat a}_{jn{\scriptscriptstyle\perp}}
        \exp
        \left \{
        {  -
        \sum_{lm}
        {{\zeta}}_{lm}(t)
        { {{\hat a}}}_{lm}
        }
        \right \}
        \over
        {\mbox{{\rm Tr}}} \,
        \exp
        \left \{
        { -
        \sum_{lm}
        {{\zeta}}_{lm}(t)
        { {{\hat a}}}_{lm}
        }
        \right \}
        }
        =
        {
        {\mbox{{\rm Tr}}} \,
        \left(
        {
        {\hat a}_{jn{\scriptscriptstyle\perp}}(\tau) -
        {\hat a}_{jn{\scriptscriptstyle\perp}}     (0)
        \over
                            \tau
        }
        \right)
        \exp
        \left \{
        {  -
        \sum_{lm}
        {{\zeta}}_{lm}(t)
        { {{\hat a}}}_{lm}
        }
        \right \}
        \over
        {\mbox{{\rm Tr}}} \,
        \exp
        \left \{
        { -
        \sum_{lm}
        {{\zeta}}_{lm}(t)
        { {{\hat a}}}_{lm}
        }
        \right \}
        }
        \end{equation}
where the contribution of higher order correlation
functions, e.g., the second term on the r.h.s. of
(\ref{4.5}),
have been neglected.
Equation (\ref{4.7}) must be considered together with
(\ref{4.3}), which refers to the constants of motion
${\hat a}_{j0}$. The parameter $\tau$ has been introduced by
the following  criteria: it is long enough to make the
dynamics of expectations
$
\langle
{\hat a}_{jn{\scriptscriptstyle\perp}}
\rangle_t
$
be independent of the history referring to times $t'<
t-\tau$; short enough to allow macroscopic state parameters
to be considered practically constant in a time interval
$\tau$ and furthermore higher order correlation functions
have been neglected to obtain (\ref{4.7}). If all this turns
out to be true, and one can safely expect that this is a
rather general situation, e.g., when the system is close
enough to the equilibrium state, then the r.h.s. of (\ref{4.7})
should be independent on $\tau$, so that one can write
        \[
        {
        {\hat a}_{jn{\scriptscriptstyle\perp}}(\tau)
        -
        {\hat a}_{jn{\scriptscriptstyle\perp}}     (0)
        \over
                            \tau
        }
        =
        {\cal L}' {\hat a}_{jn{\scriptscriptstyle\perp}},
        \qquad n \geq 1
        \]
 with ${\cal L}'$ a suitable map defined on the linear
 span of  macroscopic observables ${\hat
 a}_{jn{\scriptscriptstyle\perp}}$. By this
 map the whole dynamics can be expressed by differential
 equations generated by ${\cal L}'$:
        \[
        \langle
        {\hat a}_{jn{\scriptscriptstyle\parallel}}
        \rangle_{{\hat w}_{\zeta{(t)}}}
        =
        E_{jn{\scriptscriptstyle\parallel}}
        ;
        \qquad \qquad
        {
        d
        \over
         dt
        }
        \langle
        {\hat a}_{jn{\scriptscriptstyle\perp}}
        \rangle_{{\hat w}_{\zeta{(t)}}}
        =
        {\mbox{{\rm Tr}}} \,
        \left[
        \left(
        {\cal L}' {\hat a}_{jn{\scriptscriptstyle\perp}}
        \right)
        {{\hat w}_{\zeta{(t)}}}
        \right] .
        \]
Then one has a reduced dynamics on a time scale $\tau$,
restricted to the variables ${\hat a}_{jn{\scriptscriptstyle\perp}}$, for which the
Gibbs states ${{\hat w}_{\zeta{(t)}}}$ have the role of states.
In this way contact is established with the well-known {\it
master equation} approach to  statistical
mechanics~\cite{master-equation}: in a
sense a justification for it has been given, but with a main
difference. One does not obtain a semigroup evolution for
the statistical operator ${\hat \varrho}_t$, which at this
stage does no longer appear in the formalism, but instead
one has a dynamical map~\cite{goslar-kossakowski} ${\cal L}'$ defined on the relevant
observables, which are not constants of motion; we shall not
discuss now the problem that naturally arises of studying
expressions
$
        e^{+{{
        i
        \over
         \hbar
        }}{\hat H}t}
        {\hat a}_{jn{\scriptscriptstyle\perp}}
        e^{-{{
        i
        \over
         \hbar
        }}{\hat H}t}
$
for asymptotic times, averaged on suitable  statistical
operators, depending on the state parameters, in order to
determine the actual structure of ${\cal
L}'$. While the
Hamilton  operator plays the central role for the dynamics
described in \S~3, in this reduced dynamics description
the map ${\cal L}'$ arises, which no longer has a
Hamiltonian character~\cite{japan-goslar}.

\section{Conclusions and outlook}
Our approach points toward following developments: state
parameters $\zeta(t)$ in more general situations than
already well established hydrodynamics~\cite{Zubarev},
should be considered: e.g., a generalized chemical potential
$\mu({\mbox{{\boldmath$\xi$}}})$ related to a kinetic description
(as already pointed out in reference~\cite{Roepke}); also
the relevance of the state parameters related to the energy
dispersion, first pointed out in other
context~\cite{Ingarden} by Ingarden, should be further
investigated.
The way in which the  parameters $\zeta(t)$ and the
additional preparation  parameters determine the statistical
operator is the main point in this paper: one has the
indication that the concrete feature of the  preparation
procedure can be represented inside the formalism; so we
expect that the concept of {\it suitable preparation
procedure} should be amenable to experimental text. The
result described in \S~5 points to a dynamical semigroup
description for a restricted set of {\it slow enough}
variables~\cite{Streater}. This leads to the conjecture that some classical
insight into the microphysical dynamical behavior of the
system can be gained by means of a decomposition of the
evolution map on a suitable trajectory space. Such insight
would be desirable, since so far the classical objective
dynamics of $\zeta(t)$ arises from quantum physics for
normal modes in field theory, quite far away from classical
intuition. In much more simple settings, arising in modern
quantum optics, such a possibility has already been pointed
out~\cite{colonia-alberto}.

\end{document}